\documentclass[pre,twocolumn,superscriptaddress,showpacs]{revtex4}
\usepackage{graphicx}

\begin{document}
\title{Signatures of Dynamical Tunneling in the Wave function\\of a Soft-Walled Open Microwave Billiard}
\author{Y.-H.~Kim}
\author{U.~Kuhl}
\author{H.-J.~St\"ockmann}
\affiliation{Fachbereich Physik der Philipps-Universit\"at Marburg, Renthof 5, D-35032
Marburg, Germany}
\author{J. P. Bird}
\affiliation{Department of Electrical Engineering, the University
at Buffalo, Buffalo,  NY 14260-1920, USA}
\date{\today}

\begin{abstract}
Evidence for dynamical tunneling is observed in studies of the
transmission, and wave functions, of a soft-walled microwave cavity
resonator. In contrast to previous work, we identify the
conditions for dynamical tunneling by monitoring the evolution of
the wave function phase as a function of energy, which allows us to
detect the tunneling process even under conditions where its
expected level splitting remains irresolvable.
\end{abstract}

\maketitle

The term dynamical tunneling has been used to refer to the
phenomenon in which quantum-mechanical particles, confined within
some system, are able to access regions of phase space,
inaccessible to their classical counterparts \cite{Dav81, Tom94,
Fri98}. This process is distinct from tunneling in the usual
sense, since the "tunnel barrier" is provided by the existence of
classically-forbidden regions of phase space, rather than the
presence of any potential barrier. Recently, there has been much
interest in the manifestations of dynamical tunneling in a number
of different physical systems, such as cold atoms localized in a
superlattice potential \cite{Ste01,Hen01a}, superconducting
microwave cavities \cite{Dem00a}, and open quantum dots realized
in the high-mobility two-dimensional electron gas of semiconductor
heterojunctions \cite{Mou02, Ram03, Fer04}. Effort in these
studies has focused on demonstrating the measurable signatures of
dynamical tunneling, such as time-resolved oscillations of cold
atoms due to tunneling between symmetric KAM-island structures
\cite{Ste01,Hen01a}, tunnel splitting in the eigenspectrum of
microwave cavities \cite{Dem00a}, and periodic conductance
oscillations in quantum dots \cite{Mou02, Ram03, Fer04}. The
details of the tunneling are expected to be strongly influenced by
the structure of the classical phase space, and in Ref.
\cite{Ste01} it was found that the tunneling rate could be
strongly influenced by the presence of chaos in the corresponding
classical dynamics (chaos-assisted tunneling).

Recently, we have explored the use of open microwave cavities, as
an analog system for the investigation of transport in quantum-dot
nanostructures. Such investigations are made possible by the
equivalence, for a two-dimensional system, of the Schr\"{o}dinger and
Helmholtz equations, according to which a measurement of the
electric field within the cavity is analogous to the wave function
in the corresponding quantum dot \cite{Stoe99}. Our previous work
in this area has focused on investigations of the phenomena
exhibited by hard-walled cavities, which we have used to detect
evidence for wave function scarring \cite{Kim02}, and to
investigate the influence of a single impurity of the wave function
in quantum dots \cite{Lau94a}. In this Letter, however, we report
on the use of a novel technique to identify the existence of
dynamical tunneling, in studies of a soft-walled microwave
resonator. To the best of our knowledge, there has previously only
been one report on the implementation of a soft-walled microwave
resonator, in the unpublished thesis of Lauber \cite{Lau94a}. Such
soft-walled cavities are of particular interest, since they mimic
the form of the potential profile that is found in studies of
typical quantum dots \cite{Bir95}. The soft walls give rise to a
mixed phase space, with well-defined KAM islands that are
separated by classically impenetrable regions \cite{Fer04, Ket96}.
In previous studies of quantum dots, an analysis of the
oscillations in their magneto-conductance was argued to provide a
signature of dynamical tunneling of electrons, onto such an island
\cite{Fer04}. In this Letter, we demonstrate a novel manifestation
of dynamical tunneling in a soft-walled microwave resonator, by
studying the evolution of the phase of the wave function in the
cavity, as a function of energy (i.e. frequency). In terms of the
notation introduced by Heller \cite{Hel95}, our study reveals the
existence of "dirty" states in the wave function, which are
generated from a pair of "clean" states, degenerate in energy, are
degraded by their tunneling interaction.

In a microwave resonator with parallel top and bottom plates, the
electric field, E(x, y, z), points from top to bottom plates, the
electric field $E(x,y,z)$ points from top to bottom in the $z$
direction for the lowest modes, the transverse magnetic modes. In
the three-dimensional Helmholtz equation for $E(x,y,z)$, the $z$
dependence can be separated out, resulting in a two-dimensional
Helmholtz equation for $E(x,y)$
\begin{equation}\label{eq:helm}
\left[-\Delta_{xy}+\left(\frac{n\pi}{d}\right)^2\right]E(x,y)=k^2E(x,y)\,,
\end{equation}
where $k$ is the wave number, $d$ the height of the resonator, and
$n$ the $k_z$ quantum number. For $n=0$ this is equivalent to the
stationary Schr\"{o}dinger equation for a two-dimensional billiard
with Dirichlet boundary condition, which has been used in numerous
experiments \cite{Stoe99}. For $n=1$, however, the additional term
can be used to mimic a potential $V(x,y)$ by putting
\begin{equation}\label{eq:height}
d(x,y)\sim\frac{1}{\sqrt{V(x,y)}}\,.
\end{equation}
The approach is not exact, since the separation of the $z$ component works for constant
$d$ only, but as long as the potential variation is small on the scale of the wave
length, the error terms can be neglected.

\begin{figure}
\mbox{
\hspace{-.7cm}
\parbox[h]{0.45\columnwidth}{\vspace{-.35cm}
\includegraphics[width=0.4\columnwidth]{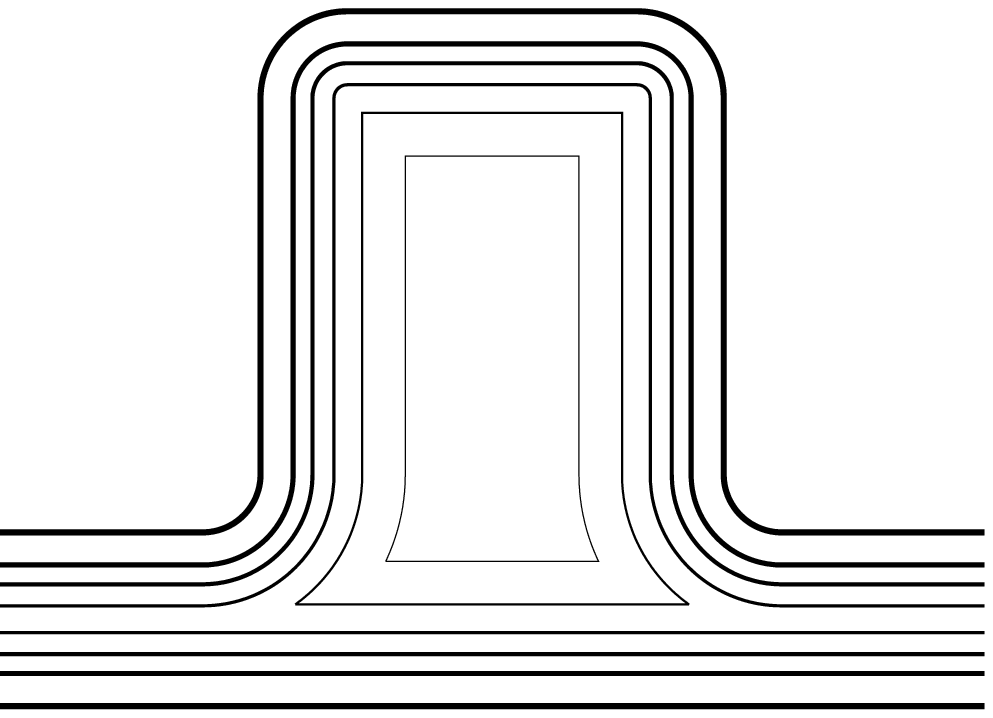}}\hfill
\parbox[h]{0.45\columnwidth}{
\includegraphics[width=0.5\columnwidth]{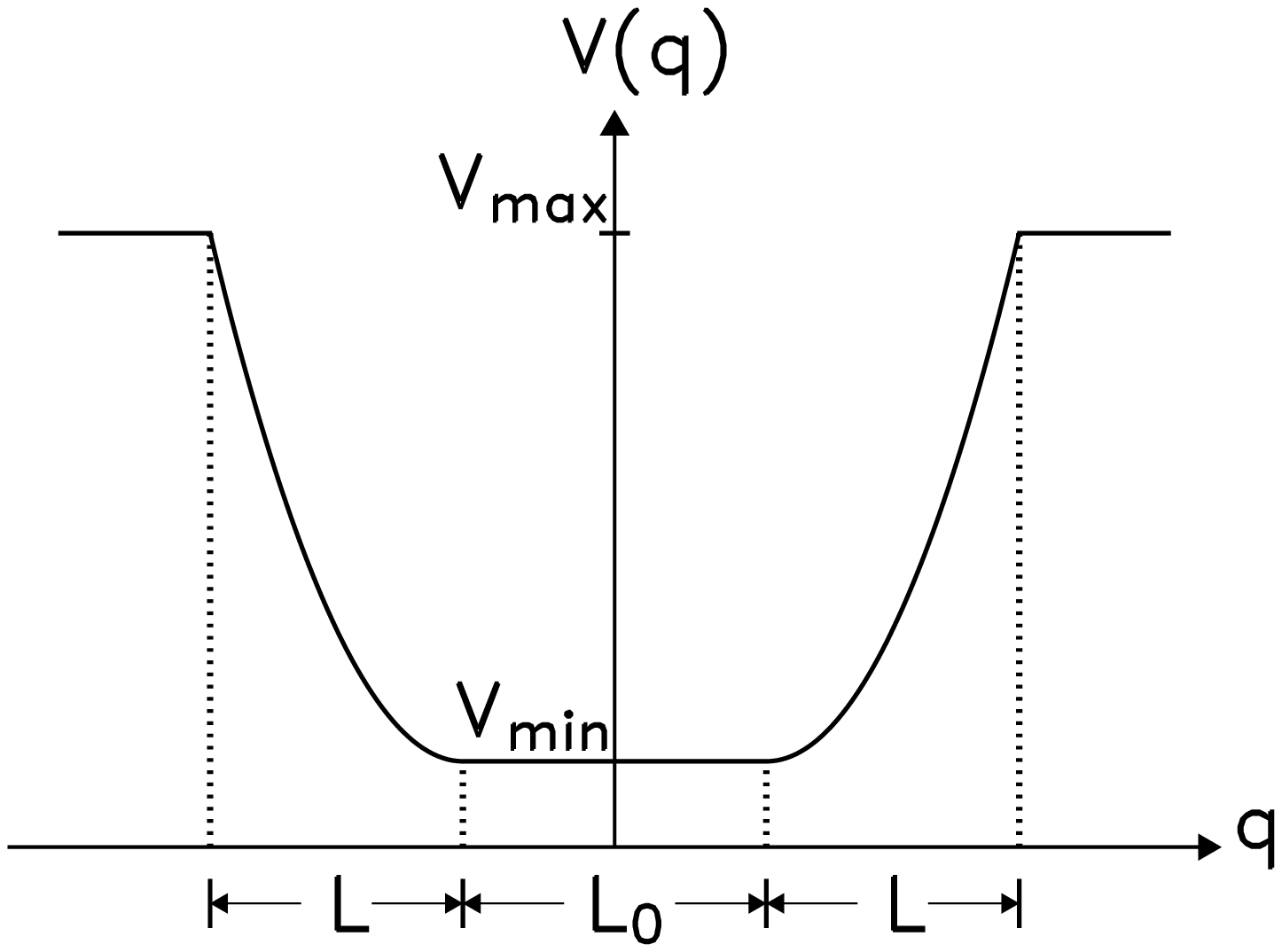}}
}
\caption{\label{fig:height}(a) Height profile of the resonator.
The top is 10.4\,mm above the bottom, between neighboring contour
lines, there is a height difference of 2.08\,mm, (b) corresponding
potential $V(q)$ along a cut, where $L=$50\,mm, and $L_0=$ 60\,mm,
140\,mm, 152\,mm for the vertical, horizontal, and diagonal cut,
respectively.}
\end{figure}

Fig. 1(a) shows a sketch of the  resonator used together with its
height profile. The minimum distance between the top and bottom
plate was $d_{\rm min}=6.3$\,mm outside the resonator increasing
gradually to $d_{\rm max}=16.7$\,mm in the bottom of the
resonator. Due to a small and unavoidable misalignment of the top
plate there were variations of the height over the area of the
resonator of up to 0.9 mm. The corresponding potential was
constant in the bottom of the dot and increased quadratically
close to the boundaries, both in the vertical and perpendicular
direction as well as along the diagonals, see Fig. 1(b). Such
potential shapes are typical for quantum dots realized by remote
surface gates \cite{Bir95}.

Two antennas $A_1$ and $A_2$ serving as source and drain for the
microwaves, were attached to the leads, which were closed by
absorbers. A third antenna $A_3$ was moved on a quadratic grid of
period 5 mm to map out the field distribution within the dot.
Details can be found in our previous publications
\cite{Kim02,Kim03c}. For $\nu_{\rm min}=c/(2d_{\rm max})=9$\,GHz
only the $n=0$ modes exist. They are not observed, however, since
the billiard is open in the $xy$ plane. For $\nu_{\rm min}< \nu
<\nu_{\rm max}=c/d_{\rm max}=$\,18\,GHz the $n=1$ modes exist as
well. Another frequency threshold of relevance is found at
$\nu_T=12.5$\,GHz. Below this frequency all $n=1$ states are bound
whereas for higher frequencies the states extend into the attached
channels and have thus to be interpreted as resonance states.

\begin{figure}
\includegraphics[width=\columnwidth]{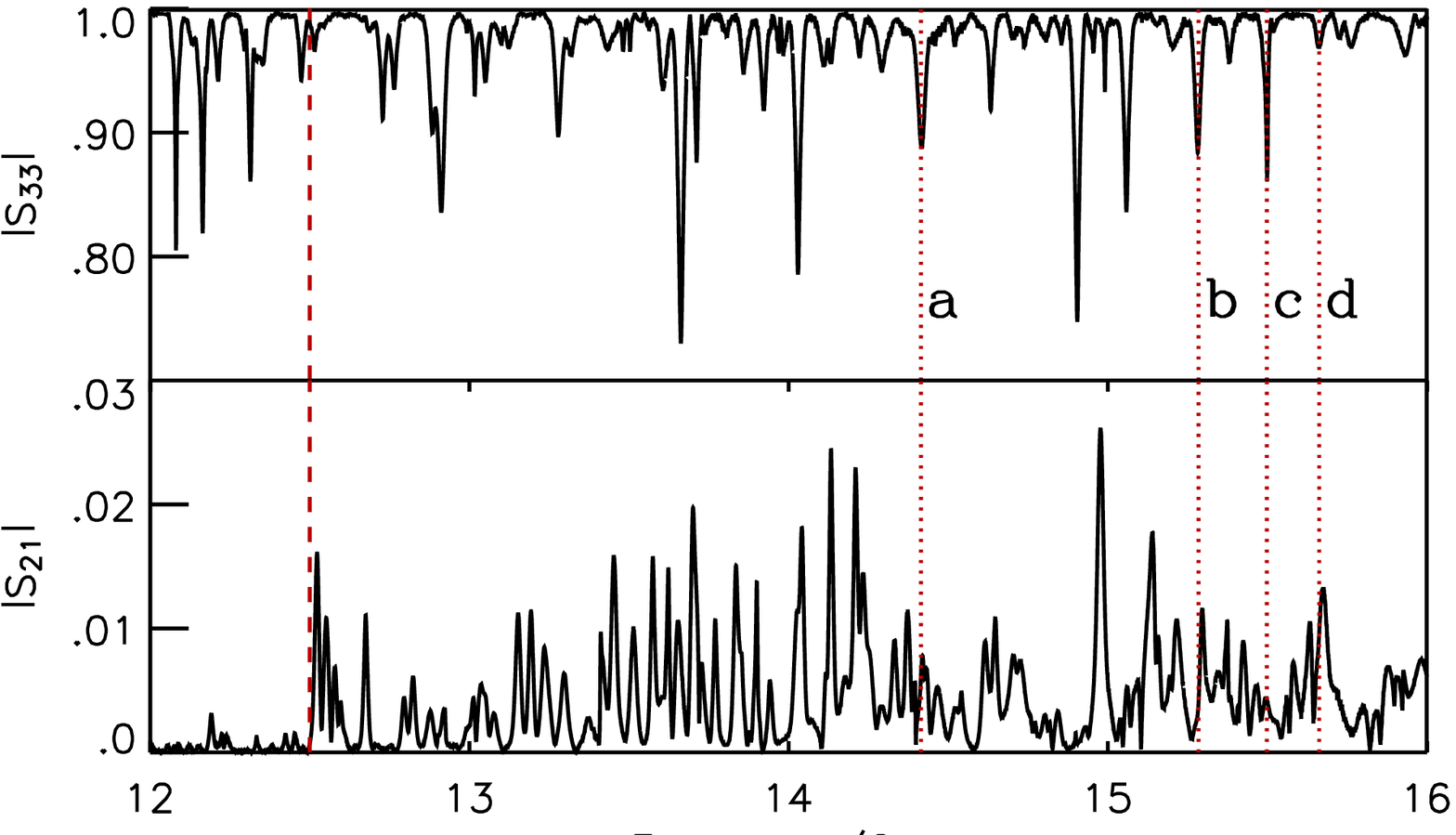}\\*[2ex]
\hspace{-.1 cm}\normalsize a:\,14.415\,GHz\hspace{1.5 cm}b:\,15.285\,GHz\hspace{1.5cm}\\
\includegraphics*[width=3.8cm]{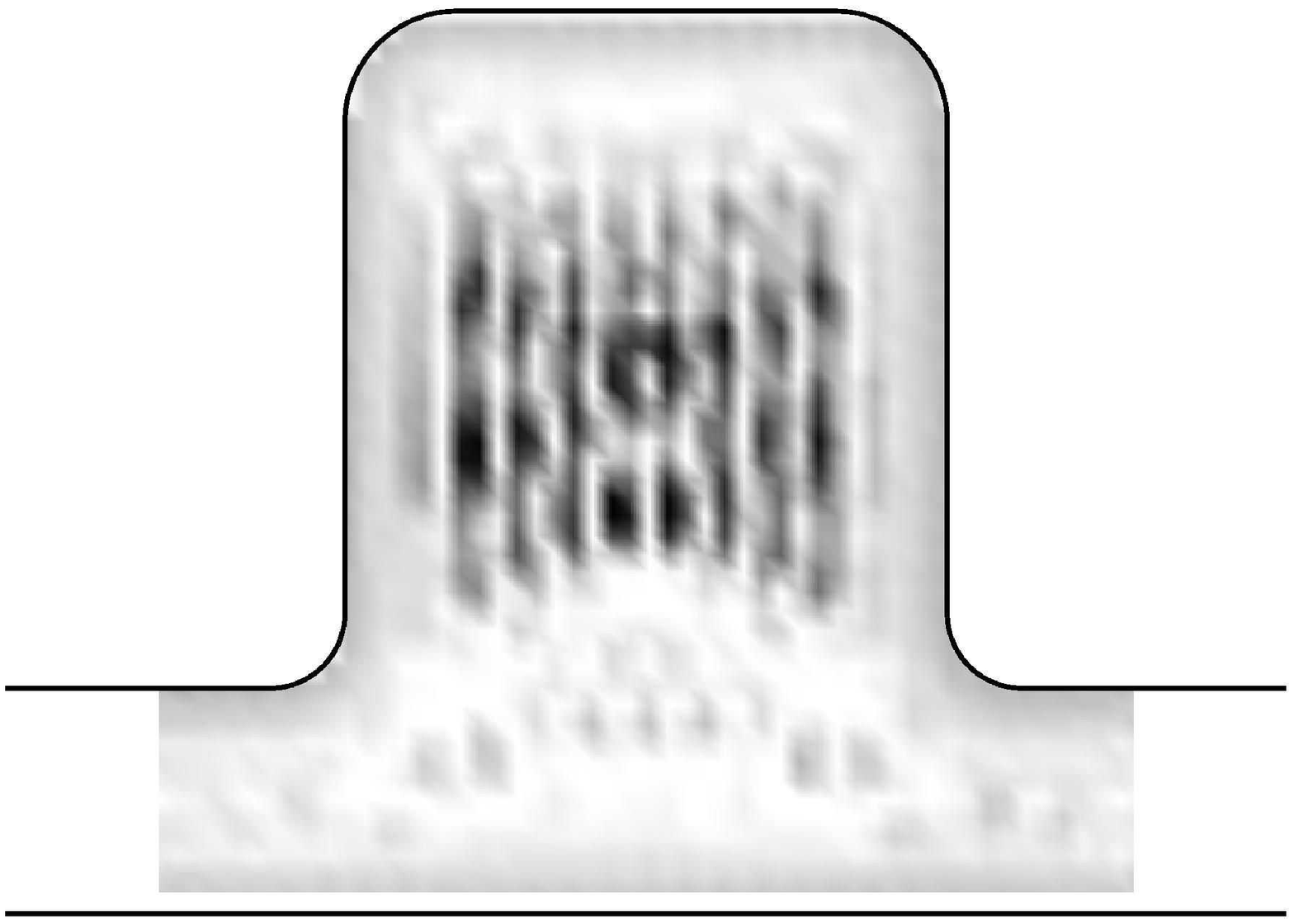}
\includegraphics*[width=3.8cm]{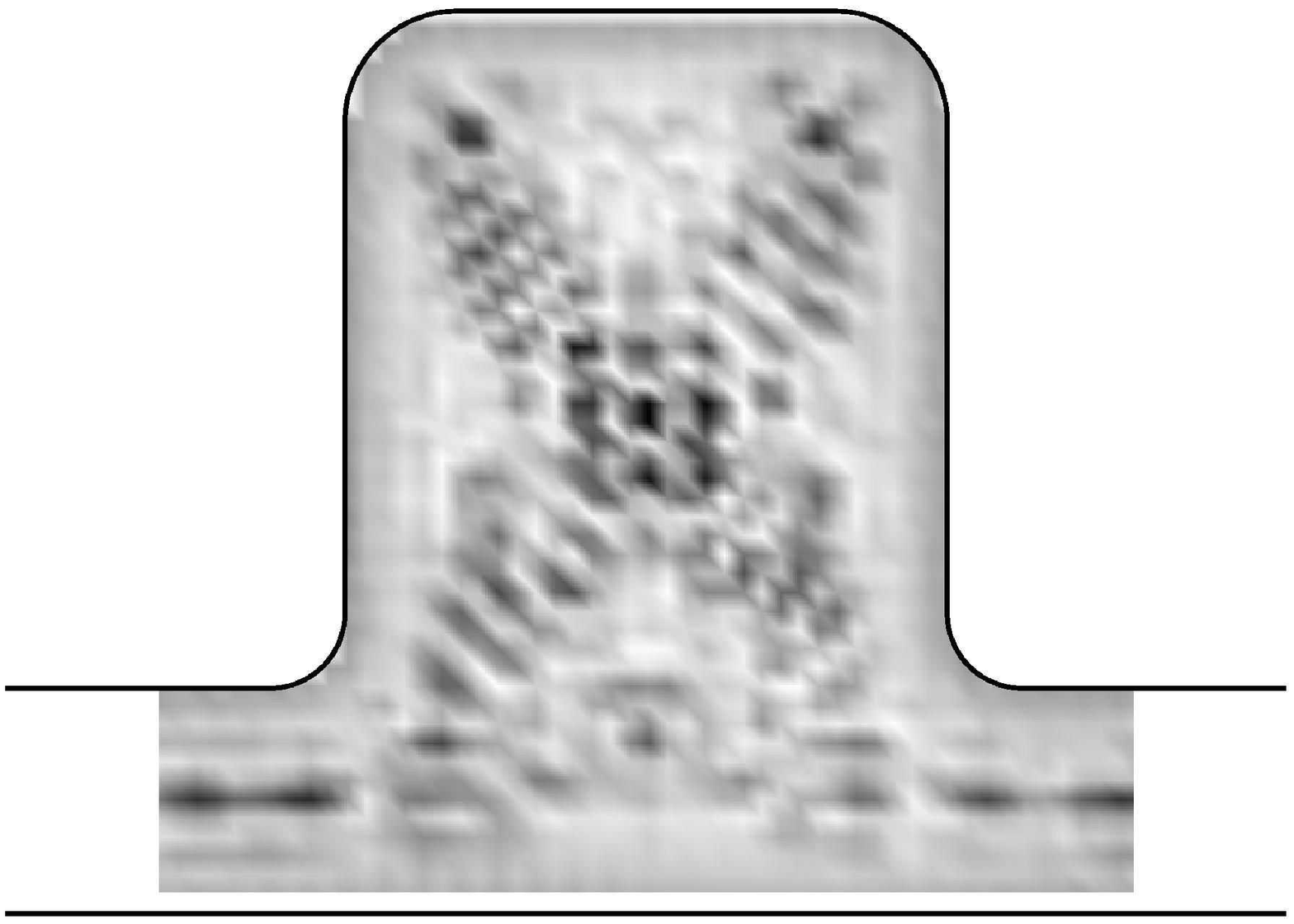}
\includegraphics*[width=3.8cm]{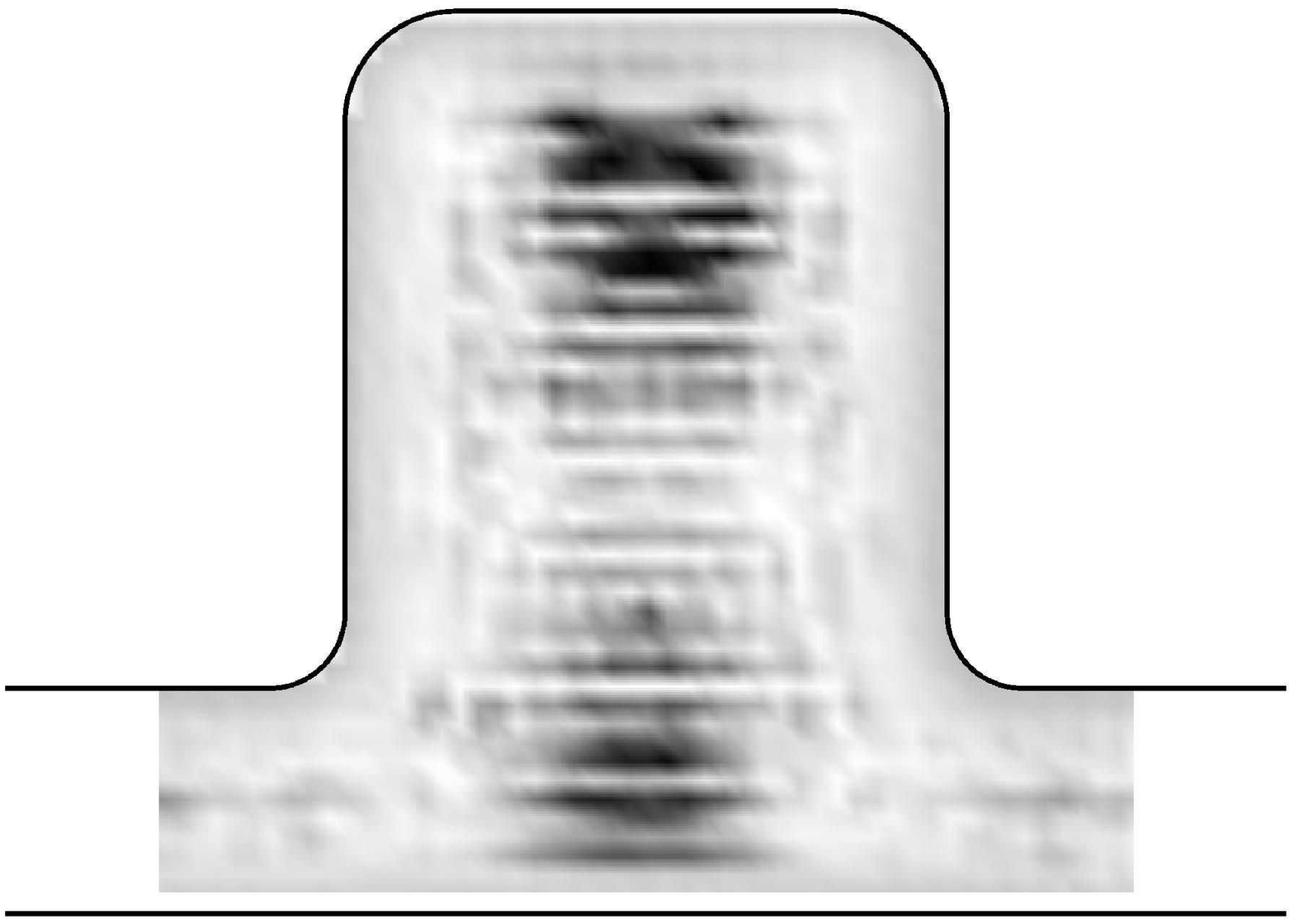}
\includegraphics*[width=3.8cm]{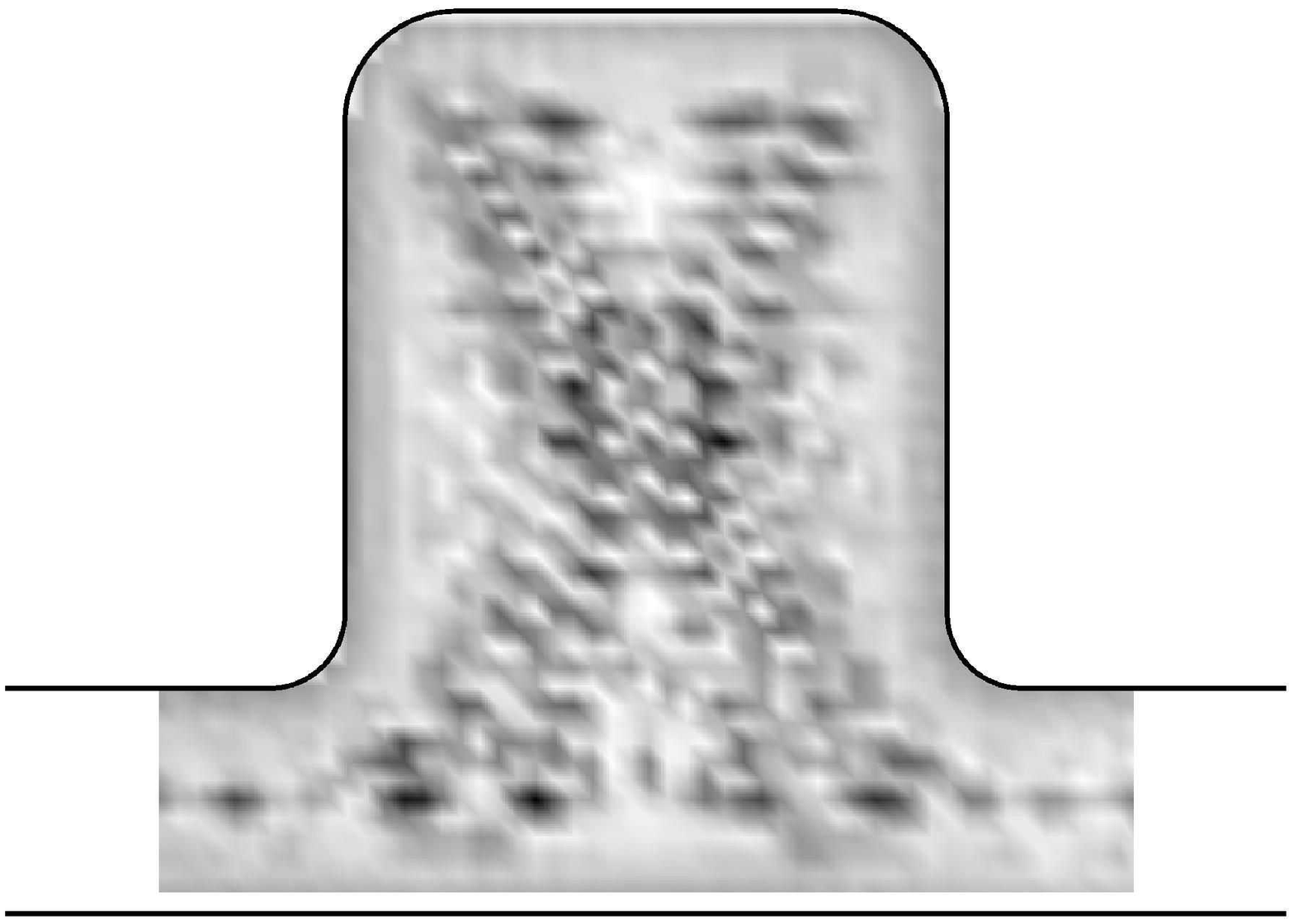}\\
\hspace{-.1 cm}\normalsize c:\,15.499\,GHz \hspace{1.5 cm}d:\,15.663\,GHz
\caption{\label{fig:spectrum}
Upper panel:
measured reflection spectrum $|S_{33}|$ with antenna $A_3$ at a fixed position,
and transmission spectrum $|S_{21}|$ between antennas $A_1$ and $A_2$
in the entrance and the exit port, respectively.
Lower panel:
wave functions for four selected frequencies obtained from the
reflection at antenna $A_3$ as a function of the position.}
\end{figure}

Fig. 2 shows the modulus of the transmission $S_{21}$ between
antenna $A_1$ in the entrance and antenna $A_2$ in the exit wave
guide, together with the modulus of the reflection $S_{33}$ at the
movable antenna $A_3$ at a fixed position within the billiard in
the upper panel. The solid vertical line at 12.5\,GHz corresponds
to the threshold frequency $\nu_T$ where the billiard states can
couple to the wave guides. Correspondingly, the transmission is
close to 0 below this frequency. The vertical dotted lines
correspond to a number of selected resonance eigenfrequencies, the
corresponding wave functions for which are shown in the lower
panel. The reflection is dominated by three scar families, two of
them associated with the vertical and the horizontal bouncing
ball, the third one corresponds to a cross-like structure. The
cross-like structure was not observed in our previous experiments
on a microwave dot with hard walls. Instead we found a scar with
the shape of a loop connecting the entrance and exit ports
\cite{Kim03c}.

A comparison of the reflection and the transmission measurements
shows that bouncing-ball states dominating the reflection
contribute only weakly to the transmission. The cross-like
structures, on the other hand,  exhibit maxima not only in the
reflection but also in the transmission spectrum. This is in
qualitative agreement with our measurements of a hard-wall
microwave dot where we also found that only those states
connecting the entrance and exit ports showed up to be relevant
for the transport.

\begin{figure}
\hspace*{0cm}\includegraphics*[width=\columnwidth]{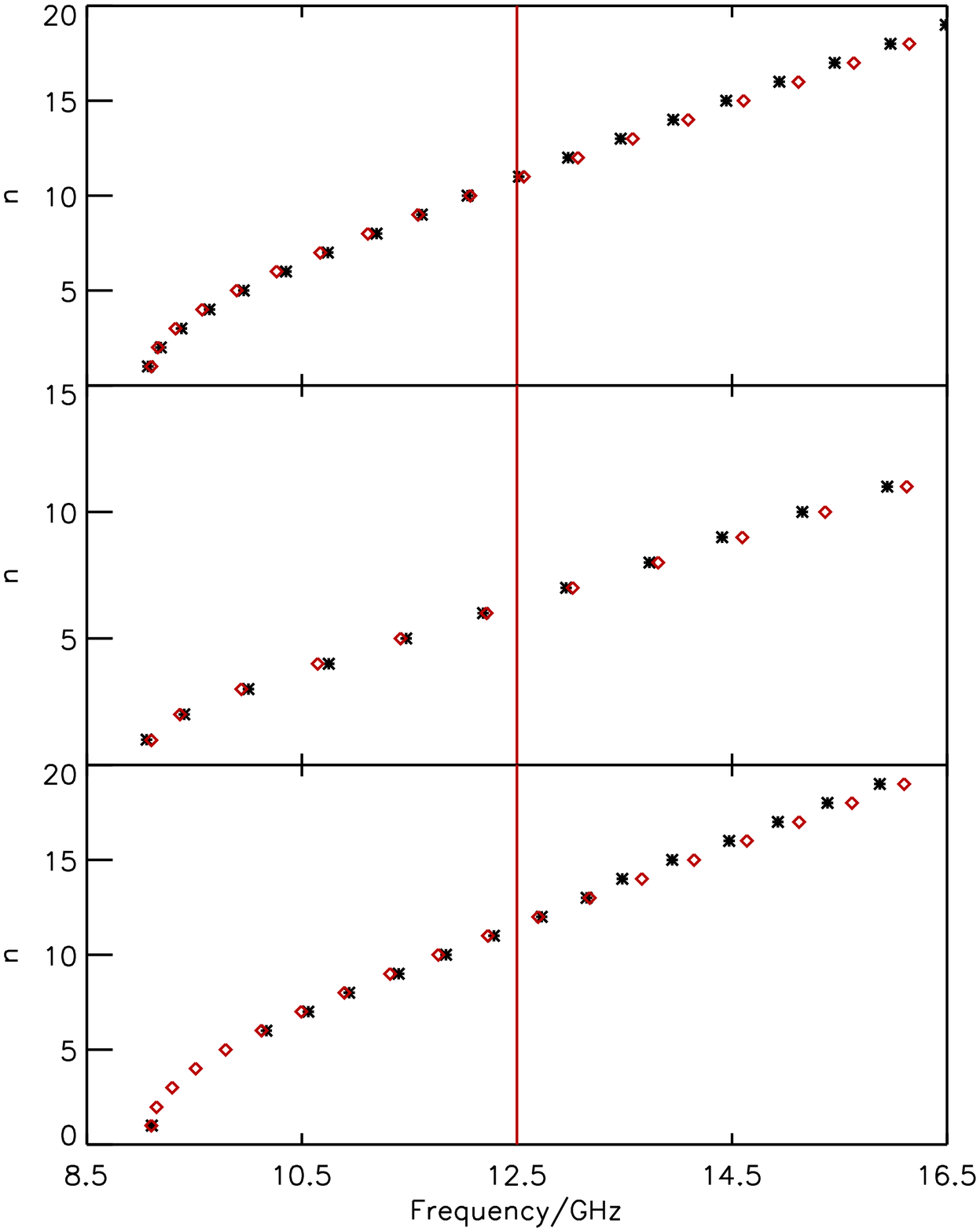}

\mbox{\raisebox{6.65cm}[0cm][0cm]{\hspace{0.25cm}
\parbox[h]{0.25\hsize}{
\includegraphics*[width=0.7\hsize]{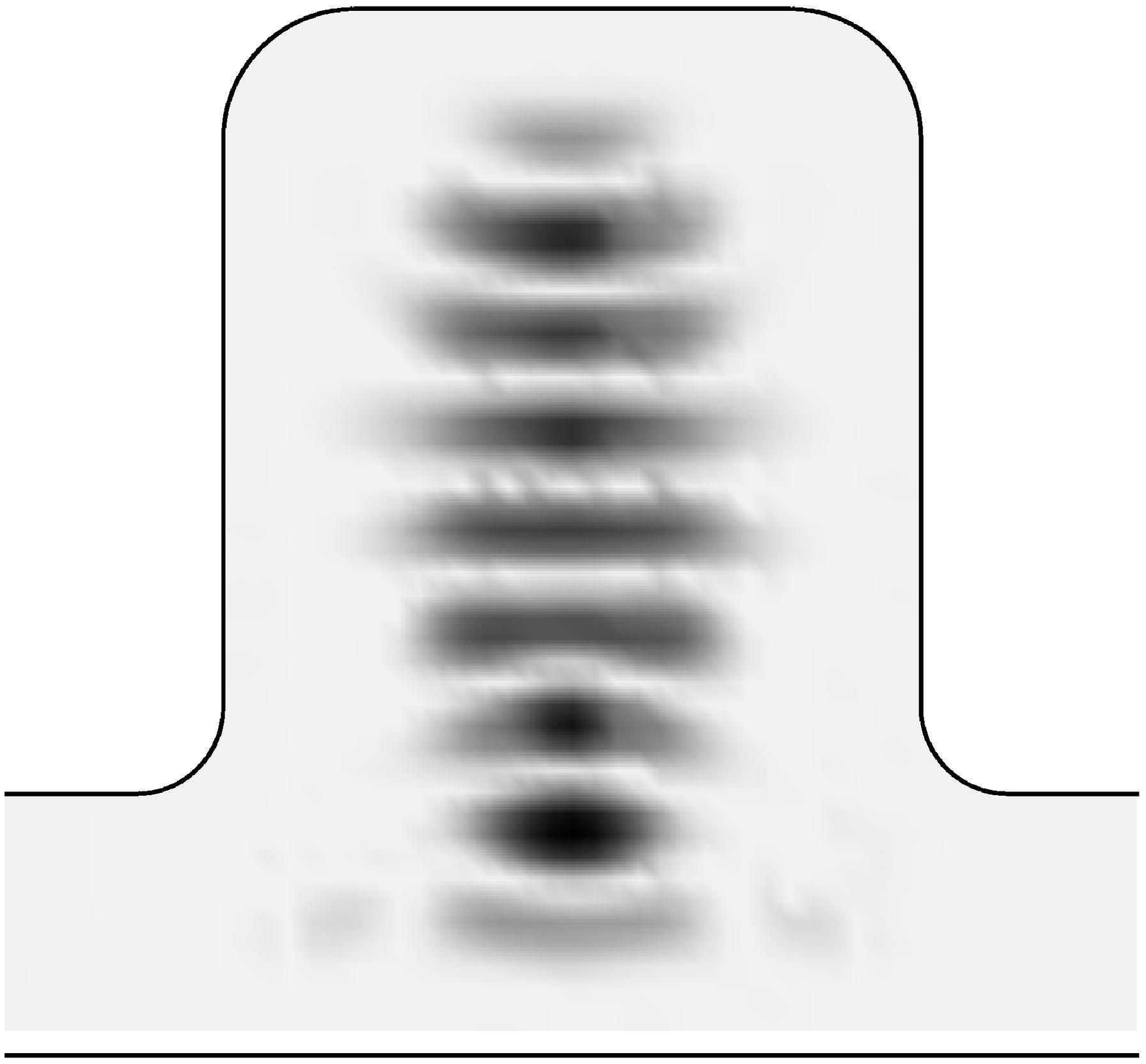}

\vspace{1.75cm}\includegraphics*[width=0.7\hsize]{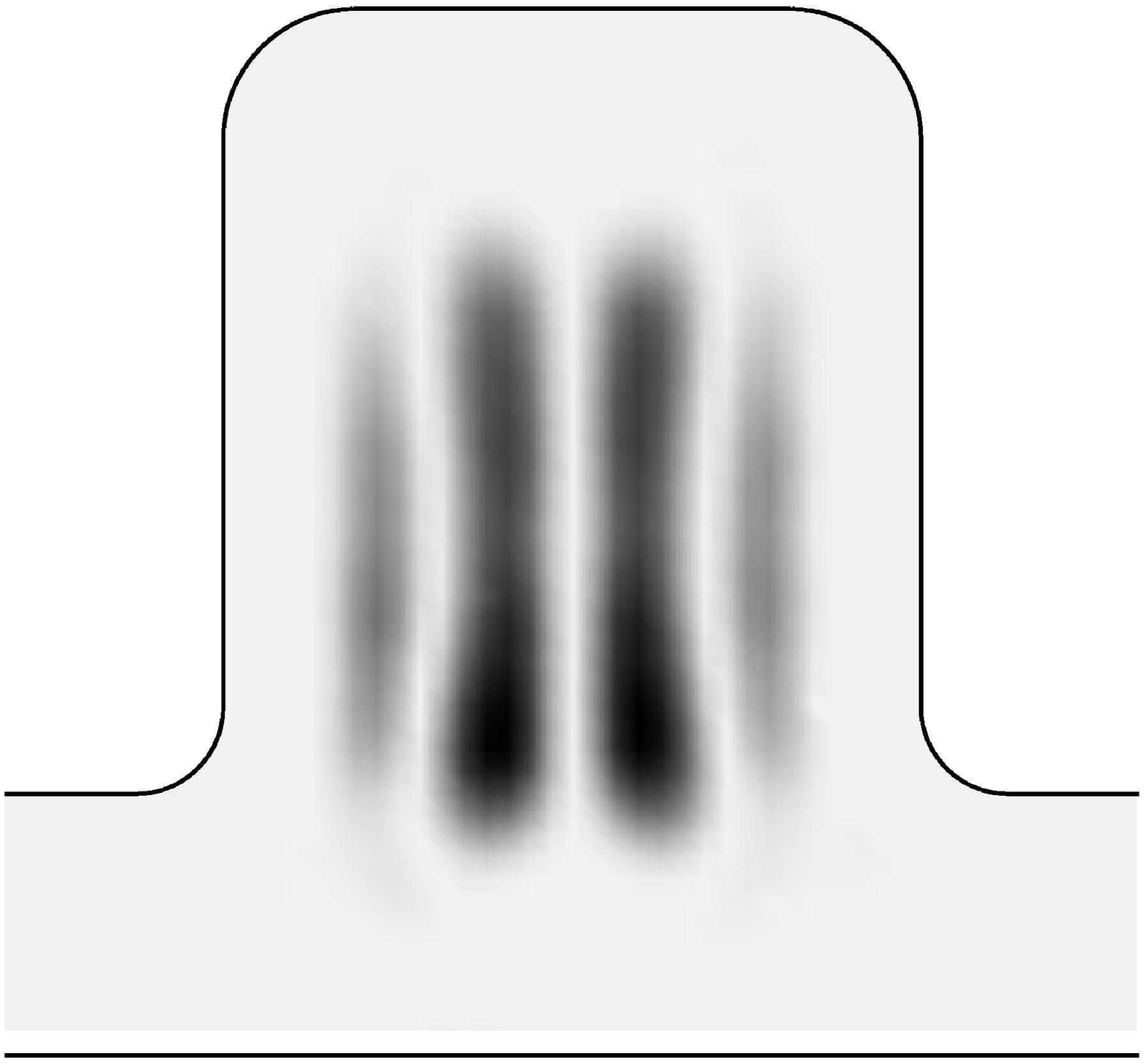}

\vspace{1.75cm}\includegraphics*[width=0.7\hsize]{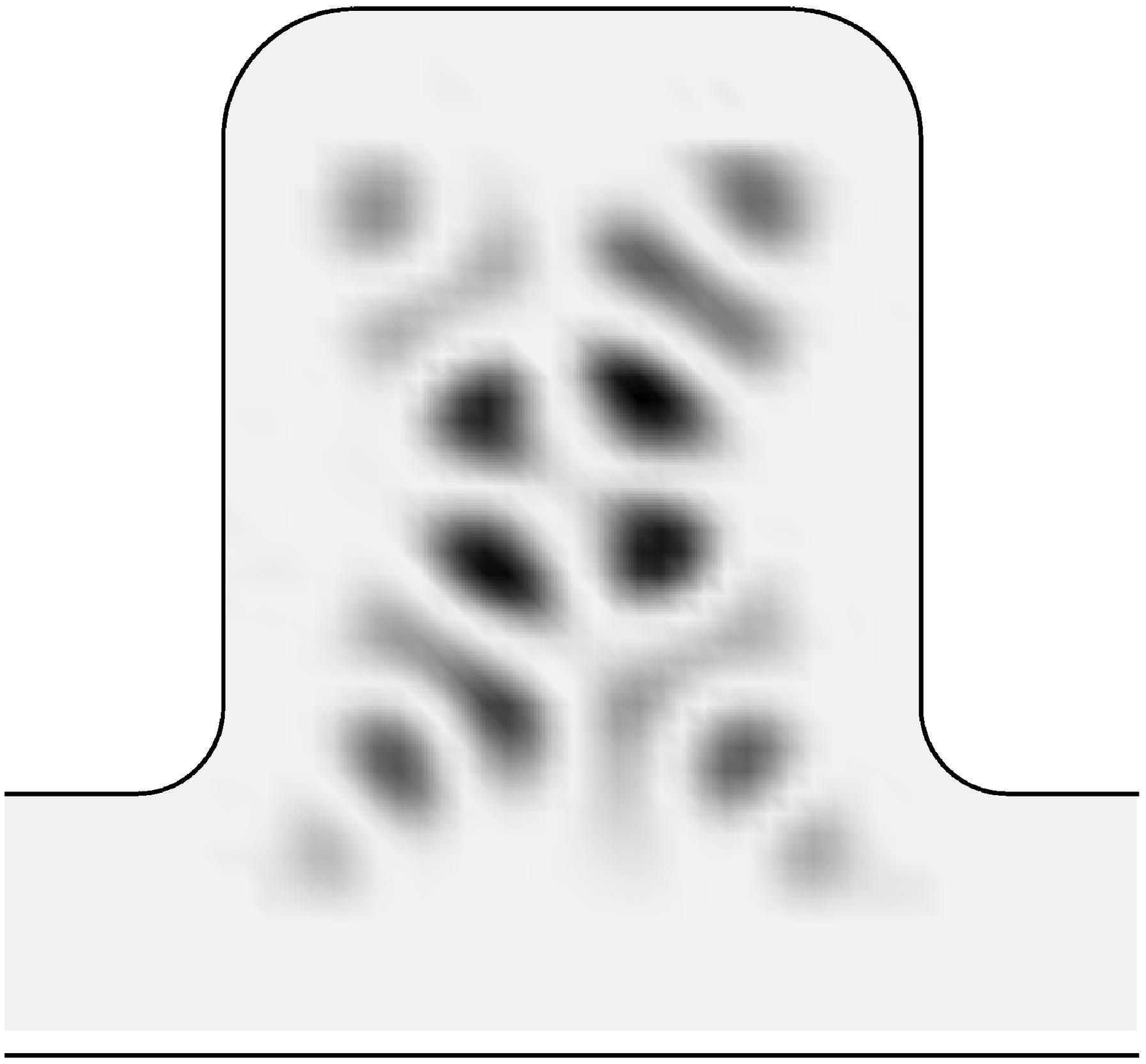}}}}
\mbox{\raisebox{5.78cm}[0cm][0cm]{\hspace{2.70cm}
\parbox[h]{0.25\hsize}{
\includegraphics*[width=0.7\hsize]{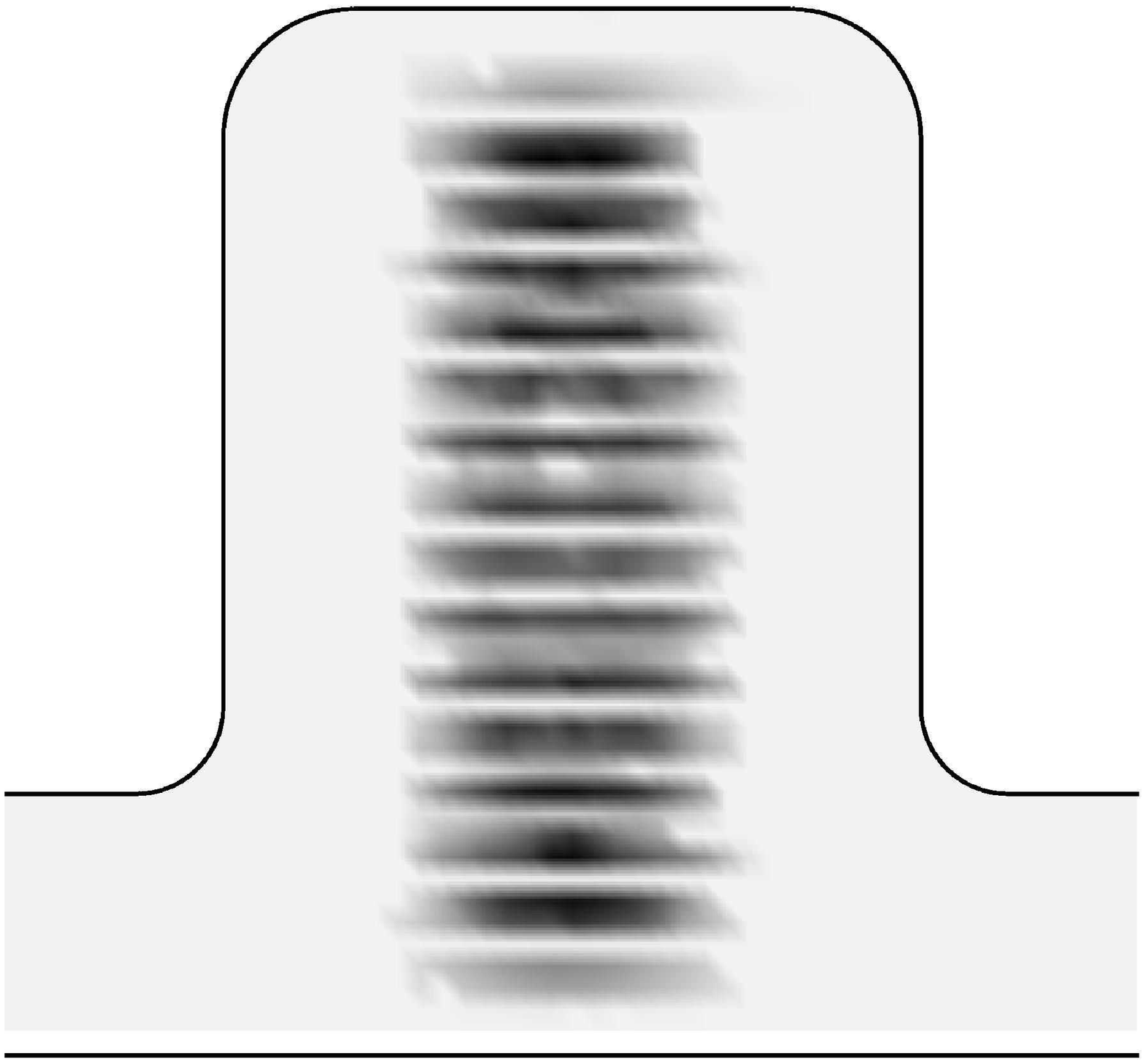}

\vspace{1.77cm}\includegraphics*[width=0.7\hsize]{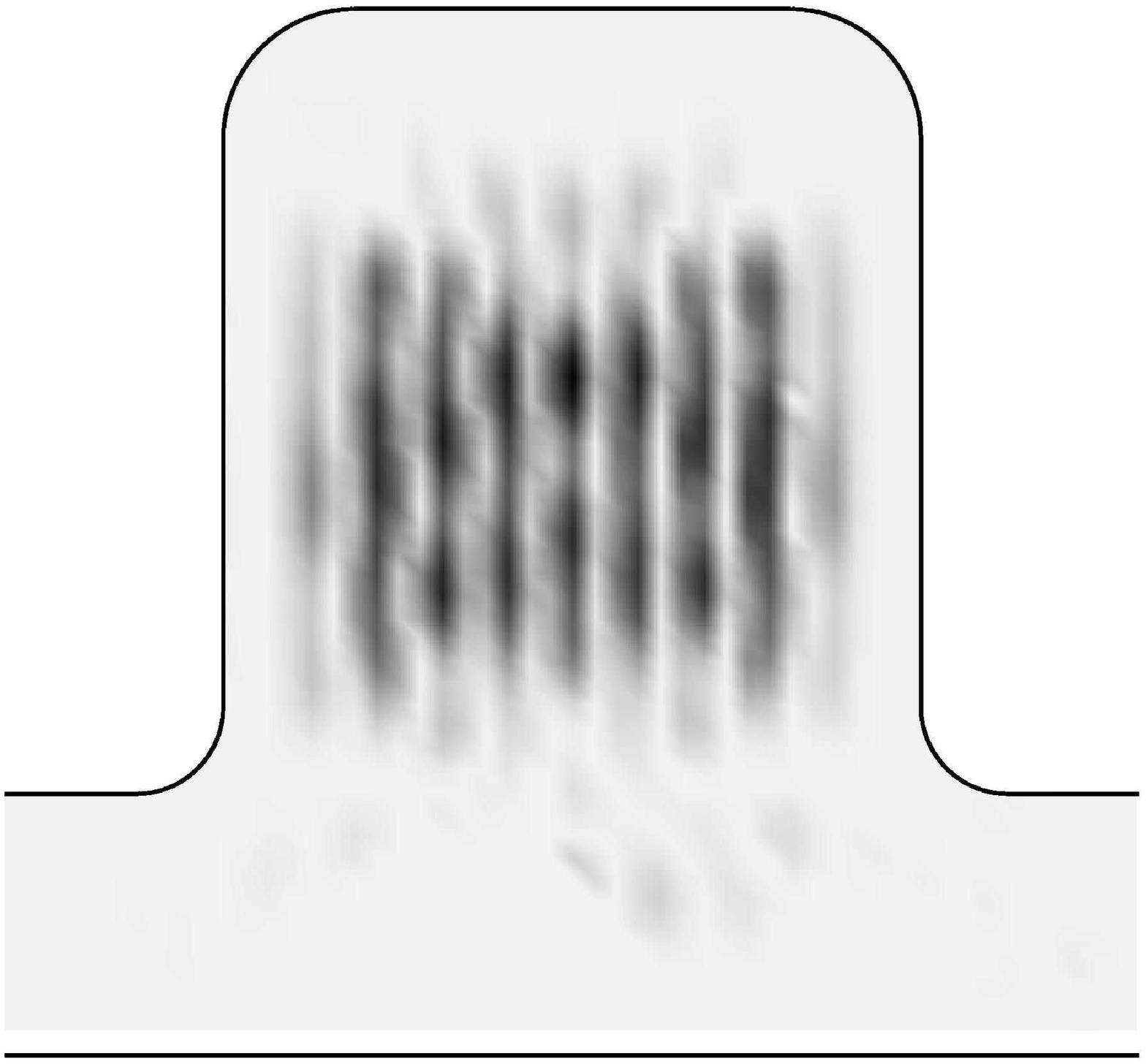}

\vspace{1.745cm}\includegraphics*[width=0.7\hsize]{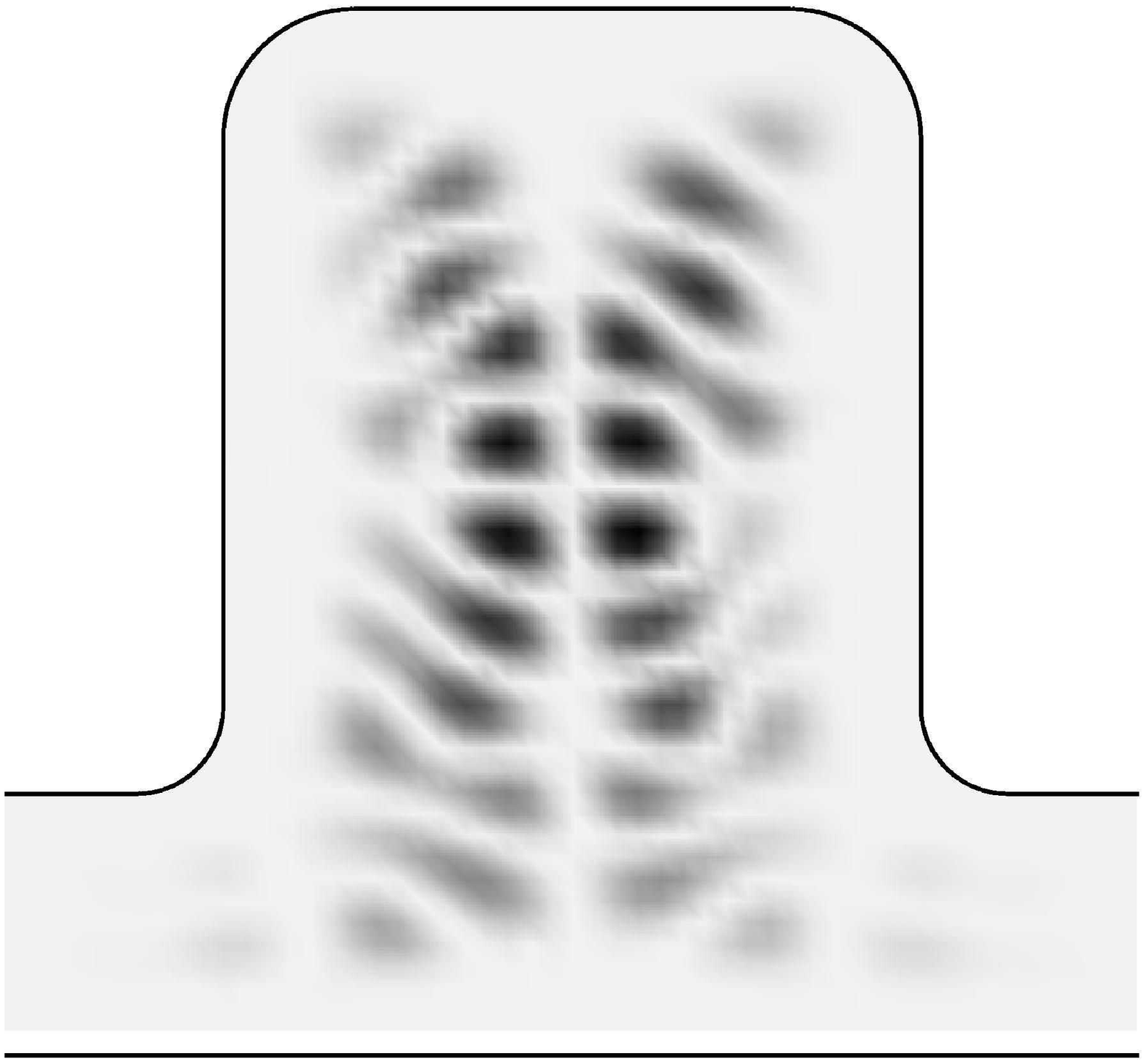}}}}

\mbox{\raisebox{7.9cm}[0cm][0cm]{\hspace{-4.77cm}
\parbox[h]{0.3\hsize}{
$n=8$

\vspace{2.88cm}$n=3$

\vspace{2.88cm}$n=7$}}}

\mbox{\raisebox{5.65cm}[0cm][0cm]{\hspace{5.4cm}
\parbox[h]{0.3\hsize}{
$n=15$

\vspace{2.88cm}$n=8$

\vspace{2.88cm}$n=12$}}} \vspace{-.5cm} \small

\vspace*{-8ex}

\caption{\label{fig:soft_boun}
Frequencies of the eigenvalues associated with the three scar families,
together with two representative members of the family.
Measured values are plotted by stars, calculated ones by diamonds.}
\end{figure}

In the first panel of Fig. \ref{fig:soft_boun} the
eigenfrequencies of all identified vertical  bouncing ball states
are shown by stars, together with two representative members of
the family. Again the vertical line denotes the threshold where
the attached channels open. The second and the third panels show
the respective results for the horizontal bouncing ball and the
cross-like scar family.

The eigenfrequencies of the members of the three families can be obtained semiclassically by
means of a WKB approximation. In one-dimensional systems with two classical turning
points $q_1,~q_2$ the action is quantized according to
\begin{eqnarray}
\label{eq:action}
S_n
&=&2\int\limits_{q_1}^{q_2}p\,{\rm d}q
=2\int\limits_{q_1}^{q_2}\sqrt{2m(E_n-V(q))}\,{\rm d}q \nonumber \\
&=&2\pi\hbar\left(n+\frac{1}{2}\right)
\end{eqnarray}
This gives an implicit expression for the eigenenergy $E_n$ of the
$n$-th state. Equation\,(\ref{eq:action}) may be applied directly
to the scarred structures observed in the experiment. For the two
bouncing-ball families a one-dimensional treatment is obviously
justified, and the cross-like structure may be looked upon as a
superposition of two one-dimensional structures, oriented along
the diagonals of the billiard. For the cross-like structures, the
application of the WKB approximation is somewhat questionable
above 12.5\,GHz where the states start to extend into the wave
guides, but this leads only to small deviations.

Inserting the potential of Fig. \ref{fig:height} into Eq.
(\ref{eq:action}), the eigenenergies for all members of the three
families can be calculated. The eigenfrequencies are given by
$\nu_n=\sqrt{2m E_n} c/(2\pi\hbar)$ and are plotted in
Fig.\,\ref{fig:soft_boun} by diamonds. The overall agreement of
the experimental results with the predicted theoretical values is
very good. To take possible misalignments between the top and
bottom plates into account, a gradient of $d_{min}$ with an
adjustable slope was allowed to improve the agreement (which was
good already without this procedure). The resulting variations of
$d_{min}$ were below the experimental uncertainties given above.

The cross-like structures deserve a separate treatment.
Semiclassically, the cross-like state is just an independent
superposition of two structures oriented along the diagonals of
the dot. Each of these structures corresponds classically to a
particle that is injected from one port, then follows the diagonal
trajectory, undergoing reflection at the upper corner, before
leaving the billiard through the same port. Classically there is
no contribution to transport. Fig.\,\ref{fig:spectrum}, on the
other hand, clearly shows that the cross-like structures do
contribute to the transport.

\begin{figure}

\includegraphics*[width=.32\hsize]{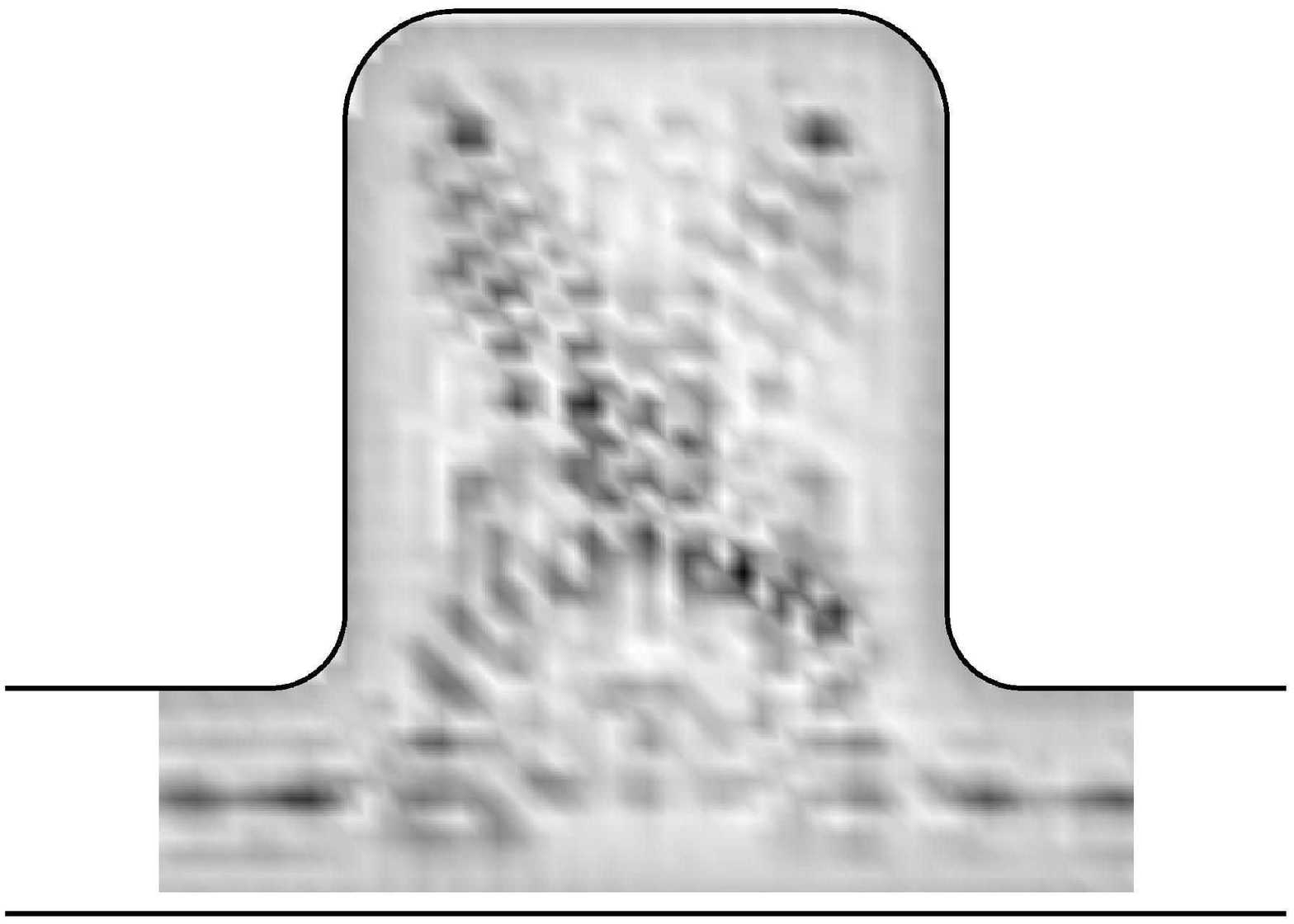}
\includegraphics*[width=.32\hsize]{fig2b}
\includegraphics*[width=.32\hsize]{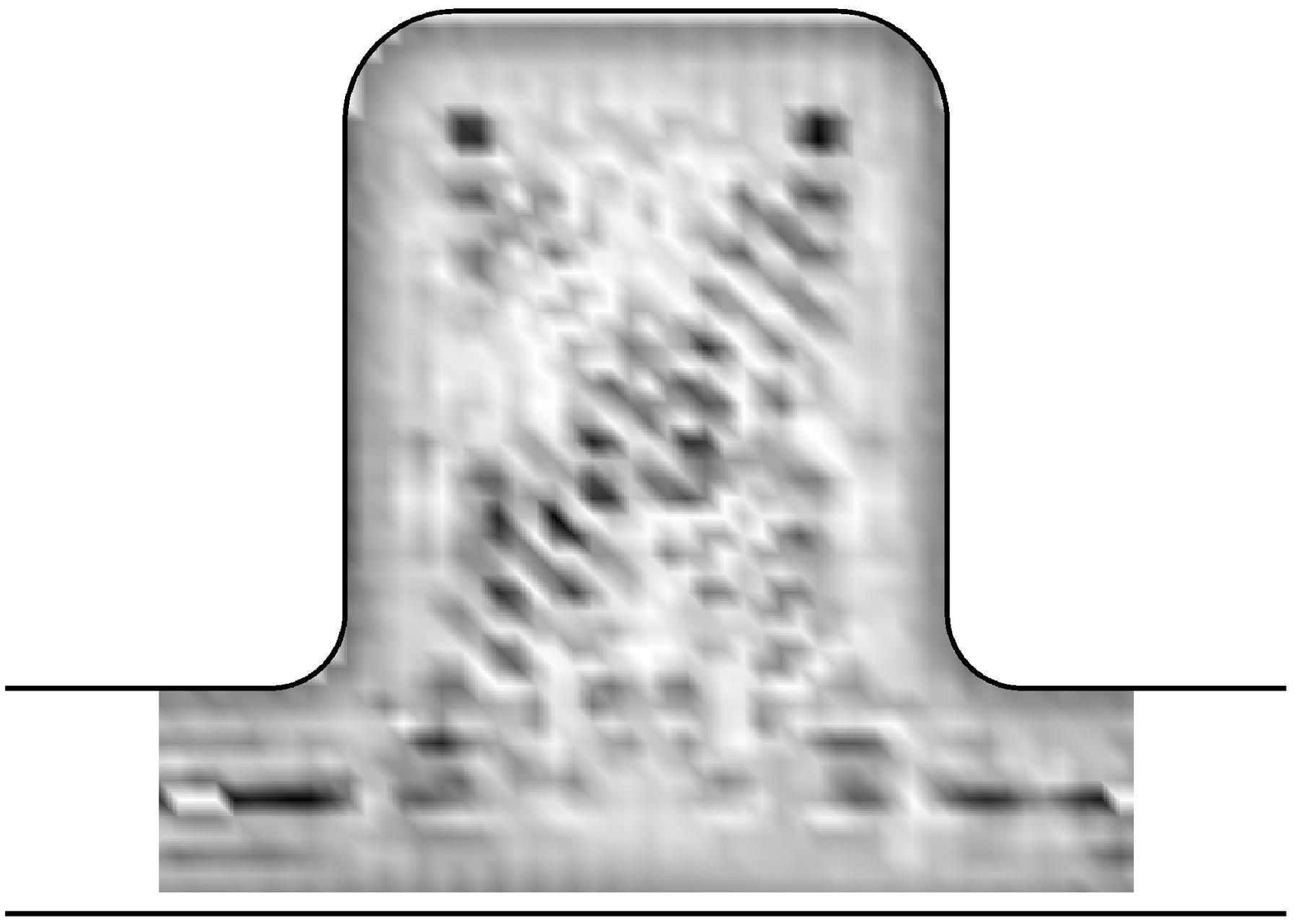}\\
\vspace{2ex}
\includegraphics*[width=.32\hsize]{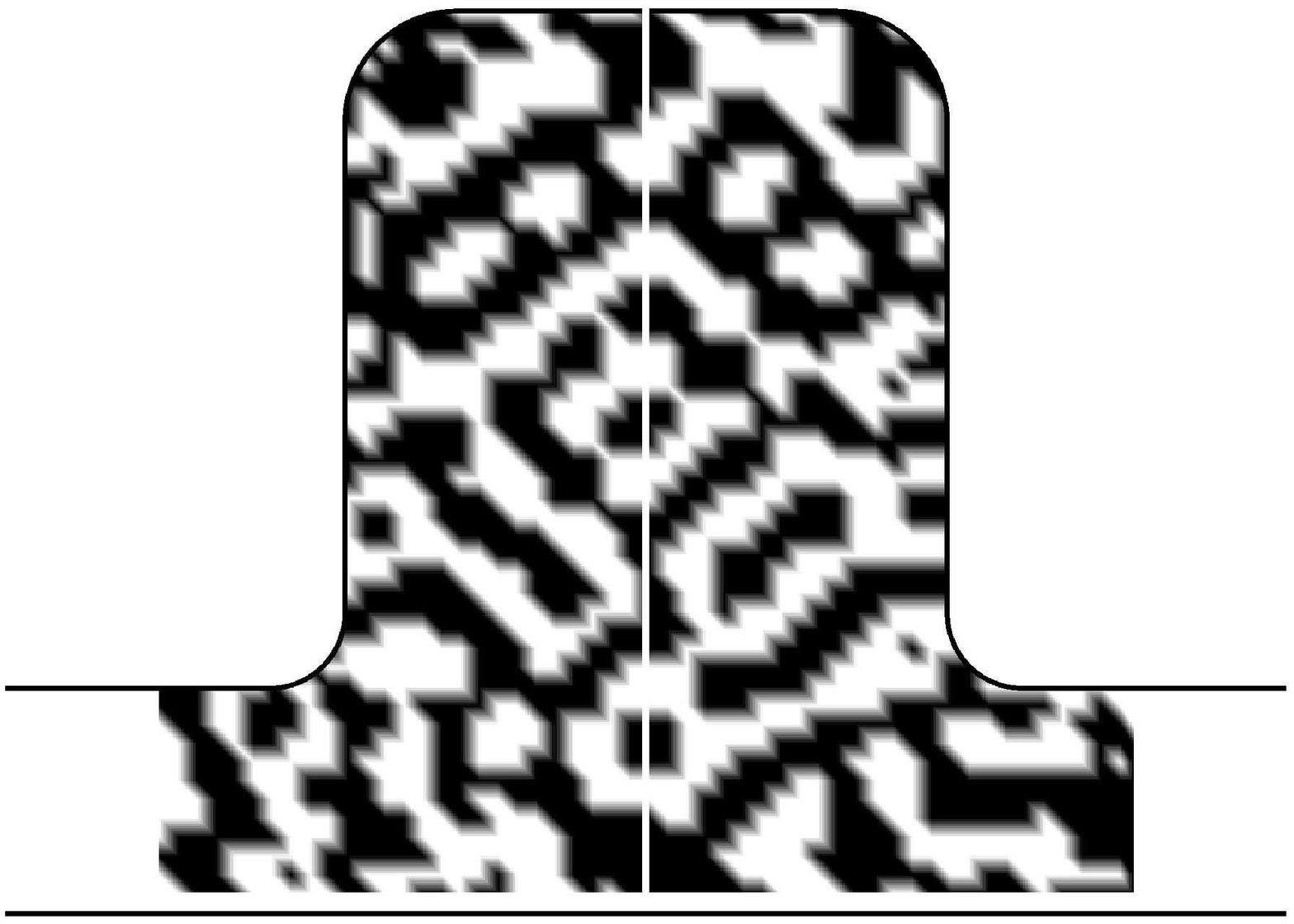}
\includegraphics*[width=.32\hsize]{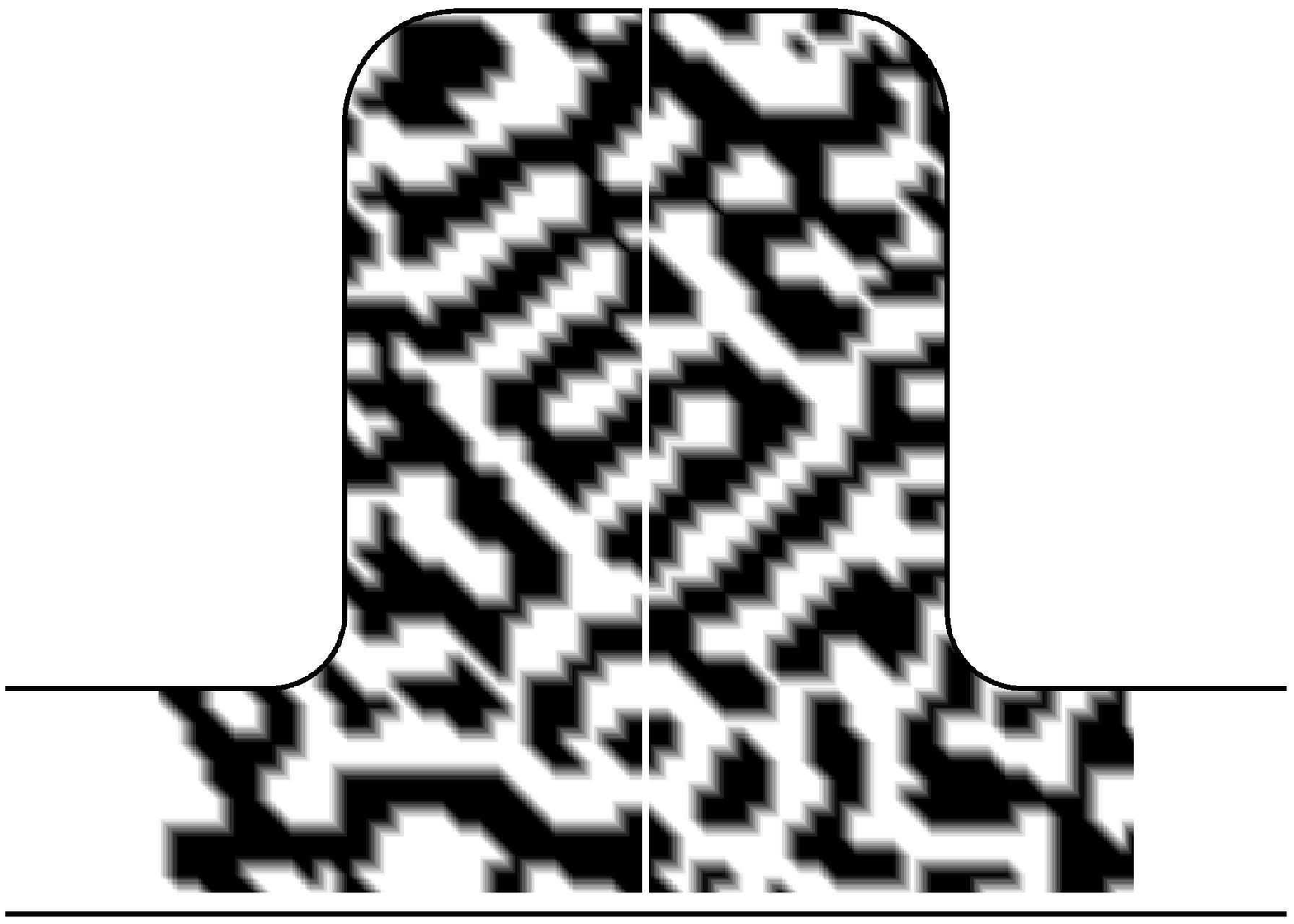}
\includegraphics*[width=.32\hsize]{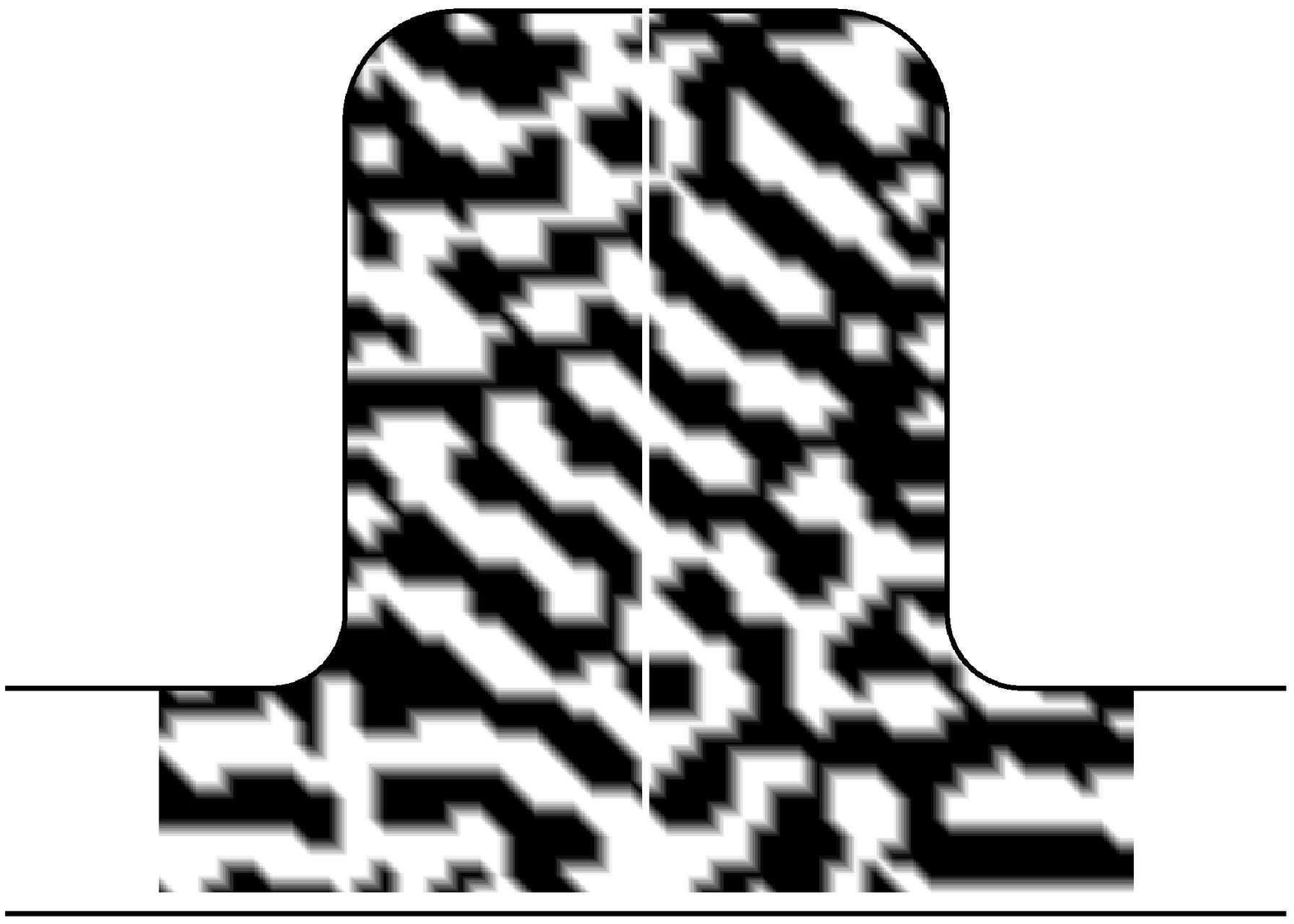}

\hspace{.0\hsize}\small
15.278GHz\hspace{.14\hsize}15.285GHz\hspace{.14\hsize}15.295GHz
\vspace{2ex}
\includegraphics[width=.99\hsize]{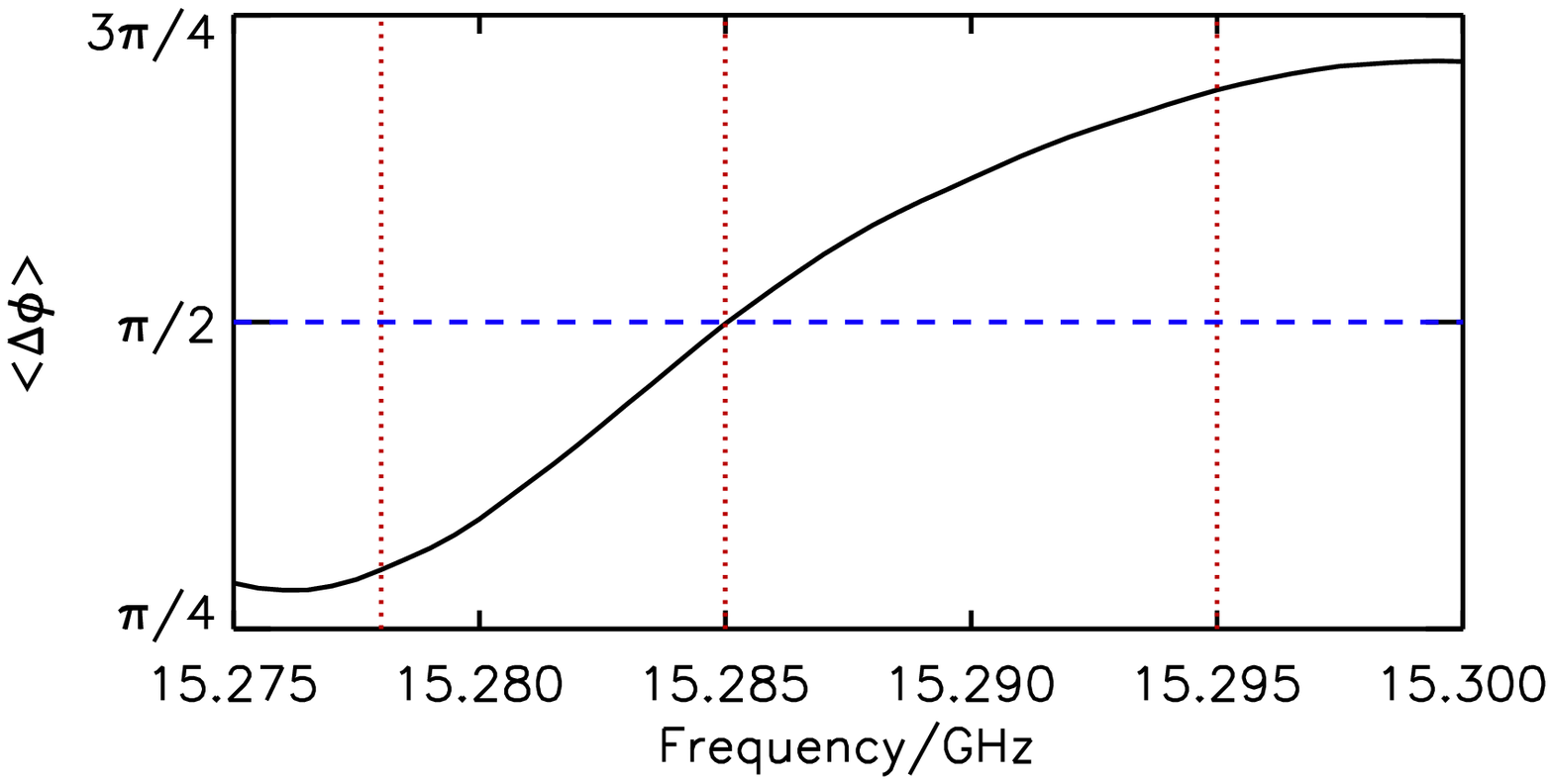}
\vspace*{-8ex}

\caption{\label{fg:phase} Upper panel: wave functions in the low
frequency wing, the center, and the high frequency wing of the
resonance, marked by the letter b in Fig.\,\ref{fig:spectrum}, as
obtained from a $|S_{33}|$ measurement. Middle panel: map of the
corresponding phases $\phi$, obtained from a $S_{31}$ measurement,
in a black-white plot. Bottom part: phase asymmetry
$\langle\Delta\phi\rangle$ as a function of frequency.}
\end{figure}

So there must be a quantum-mechanical admixture of the states, or,
expressed in other words, a dynamical tunneling coupling of these
states. This, however, implies that the originally two-fold
degenerate states split into doublets, with a symmetric wave
function associated with the lower, and an anti-symmetric one with
the higher energy. An inspection of the spectrum unfortunately
does not show any indication of a doublet splitting (see, e.\,g.
the resonances marked by the letters b, d in Fig.
\ref{fig:spectrum}). This could not be expected, anyway. In the
only previous microwave experiment on chaos-assisted tunneling,
the observed splittings were below 1\,MHz \cite{Dem00a}, much
smaller than the line widths observed in the present set-up.
Superconducting resonators were essential to resolve such
splittings. In the open microwave dot used in this work,
superconducting cavities would not have been of use anyway, since
the line widths are limited mainly by the openings, and not by the
absorption in the walls.

\begin{figure}
\includegraphics*[width=.99\hsize]{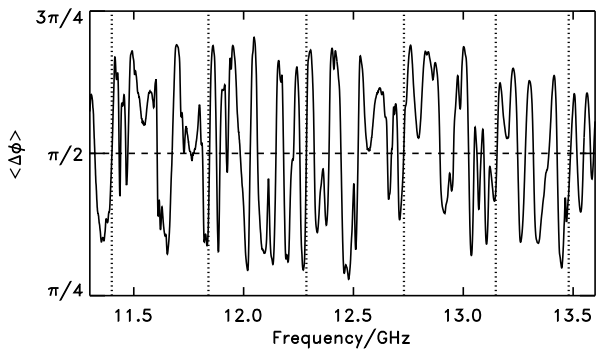}
\caption{\label{fg:phase1} Phase asymmetry
$\langle\Delta\phi\rangle$ as a function of frequency. The
frequencies corresponding to cross-like wave functions are marked
by vertical dashed lines.}
\end{figure}

Fortunately there is an alternative means to obtain  direct
evidence of dynamical tunneling, even in cases where the line
splitting cannot be resolved. This is illustrated in Fig.
\ref{fg:phase}. The upper panel shows again the wave function at
15.285\, GHz, marked by the letter b in Fig. \ref{fig:spectrum},
but in addition the wave functions in the lower and higher
frequency wing of the resonance are shown as well. There is no
noticeable difference between the three patterns. A completely
different impression emerges, however, when, in addition, the
phases obtained from the transmission $S_{31}$ are also
considered. In the middle panel of Fig. \ref{fg:phase} the
corresponding phase maps are depicted, demonstrating without any
doubt that the wave function is symmetric in the low, and
anti-symmetric in the high frequency wing of the resonance. To
make this even more evident, we calculated the phase asymmetry of
the wave function via
\begin{equation}\label{eq:phase}
\langle\Delta\phi\rangle=\langle\phi(x,y)-\phi(-x,y)\rangle\,,
\end{equation}
where the average is over the area of the dot.
$\langle\Delta\phi\rangle$ should be 0 for the symmetric case and
$\pi$ for the antisymmetric case. The bottom part of Fig.
\ref{fg:phase} shows the phase asymmetry for the 15.285\,GHz
resonance as a function of frequency. Though the ideal values 0
and $\pi$ are not obtained, a change from symmetric to
anti-symmetric behavior while passing through the resonance is
unmistakable. Fig. \ref{fg:phase1} shows the phase asymmetry over
a larger frequency range. Frequencies associated with cross-like
resonances are marked by vertical dotted lines. For each of these
frequencies the phase asymmetry passes through  $\pi/2$, from
below to above with increasing frequency. We have thus obtained a
direct evidence of dynamical tunneling, using nothing but the
change of the symmetry properties of the wave function upon
passing through the resonances.

In conclusion, we have demonstrated a novel manifestation of
dynamical tunneling in a soft-walled microwave resonator. The wave
function of this system exhibits scarring due to a number of
different bouncing orbits, and the eigenfrequencies of these scars
were shown to be well described by the WKB approximation. In
contrast to previous work, where dynamical tunneling has been
identified by detecting its associated splitting of the
eigenspectrum, in this report we obtained direct evidence for the
tunneling process by studying the evolution of the wave function
phase as a function of energy (i.e. frequency). This allowed us to
identify the conditions for dynamical tunneling, even though its
related level splittings were irresolvable in this system.

This work was supported by the Deutsche Forschungsgemeinschaft via
individual grants. Work at the University at Buffalo is supported
by the Department of Energy, the Office of Naval Research, and the
New York State Office of Science, Technology and Academic Research
(NYSTAR)

\bibliography{thesis,paperdef,paper,newpaper,book}

\end{document}